\documentclass[12pt]{revtex4}

\begin{document}

\title{Physical meaning of Lagrange multipliers}

\author{Hasan Karabulut}

\address {Rize University,\\
Faculty of Arts and Sciences, Physics department,\\
53100 Rize/TURKEY}

\begin{abstract}
A rule to assign physical meaning to Lagrange multipliers is
discussed. Examples from mechanics, statistical mechanics and
quantum mechanics are given.
\end{abstract}

\maketitle

\section{Introduction}

Lagrange multipliers arise frequently in physics, engineering, economics and
mathematics in optimization problems with constraints. In statistical
mechanics it arises in microcanonical derivation of distribution laws of
quantum gases and in some other problems where entropy or free energy is
maximized or minimized under the constrains of fixed total particle number
and/or energy. In classical mechanics Lagrange multipliers arise in
Lagrangian formulation of mechanics with constraints where multipliers are
interpreted as constraint forces. In economics and engineering there are
plenty of problems that requires optimization of some quantity under some
constraints and Lagrange multipliers arise naturally in these fields.

In most of these cases Lagrange multiplier turns out to have a physical
meaning at the end of calculation. In economics people seem to have long
been aware that the multipliers are related to rate of change of optimized
quantity with respect to some parameter in the calculation. In physics it is
well known that multipliers usually have precise physical meanings although
there is not a general rule to identify their physical meaning directly. In
statistical mechanics Lagrange multipliers occurs in many places and the
pedagogical articles on them seem to concentrate on how to teach statistical
mechanics without using Lagrange Multipliers\cite{1a,1b,1c,1d,1e} rather
than how to find precise physical interpretation of the multipliers.
Statistical mechanics textbooks derive physical meaning of multipliers by
using supplementary arguments. I have long felt that (at least in
statistical mechanics) there must be a direct rule to identify the physical
meaning of multipliers and after some research I found that such a rule
exist and It works well if the constrains have a well defined physical
meaning. It does not seem to work for quantum applications in particular
where quantities such as overlap integrals between orbitals do not have a
clear physical meaning. But in some areas of physics and engineering as well
as in economics the rule works well.

I have come across this rule in economics textbooks and economists have been
aware of it for long time. Even the books on mathematical economics written
in the sixties and seventies have the rule\cite{2} and there are
articles(see Baxley and Moohouse\cite{3} for example) discussing the
Lagrange multipliers in economics. Engineers too are interested in Lagrange
multipliers and Bertsekas's book\cite{4} on Lagrange multipliers has the
above mentioned rule. Traditionally Lagrange multipliers method is
introduced in calculus books and they do not discuss physical meaning of
multipliers. Even this is changing and two of recent calculus books\cite
{5a,5b} discuss the meaning of multipliers and present the rule without
proof(Actually \cite{5b} provides some justification). The Lagrange
multipliers are used in constrained variational problems and a proper
extension of the rule exists in the literature on calculus of variations.%
\cite{6}

I have checked the statistical mechanics books that I was able to reach and
none of them mentions the rule and I have not come across it elsewhere in
the physics literature that I am familiar with. It seems that the rule is
not well known in physics community. This prompted me to write this paper
and discuss use of this rule in physics. In the paper I will show how the
rule makes interpretation of multipliers easy in statistical mechanics and I
will also present a quantum application where the rule is not applicable. I
also provide some examples that shows how the rule makes interpretation of
Lagrange multipliers easy in some mechanics problems.

In the following section I derive the above mentioned rule for discrete
variables and I discuss two examples. In the third section I derive the same
rule for constrained variational problems and I discuss three examples.

\section{ Derivation of the rule for discrete variables and examples}

First we start by reminding the reader how Lagrange multipliers used in
constrained optimization. Suppose we want to find a local maximum or minimum
of a function $f(x_{1,}x_{2},...,x_{N})$ under the constraints
\begin{equation}
\begin{array}{ll}
g_{i}(x_{1,}x_{2},...,x_{N})=u_{i}, & (i=1,2,...,M).
\end{array}
\label{a20}
\end{equation}
Lets denote the maximum/minimum point with starred characters $%
(x_{1,}^{*}x_{2}^{*},...,x_{N}^{*})$ and lets use the compact vector
notation $\overrightarrow{r}=(x_{1,}x_{2},...,x_{N})$ and $\overrightarrow{r}%
^{*}=(x_{1,}^{*}x_{2}^{*},...,x_{N}^{*})$. We also use N-dimensional
gradient
\begin{equation}
\overrightarrow{\nabla }f=\frac{\partial f}{\partial x_{1}}\widehat{e}_{1}+%
\frac{\partial f}{\partial x_{2}}\widehat{e}_{2}+...+\frac{\partial f}{%
\partial x_{N}}\widehat{e}_{N},  \label{a10}
\end{equation}
where $\widehat{e}_{1},\widehat{e}_{2}...\widehat{e}_{N}$ are the unit
vectors in the directions $x_{1,}x_{2},...,x_{N}$. With this notation, at
the extremum point we must have

\begin{equation}
\overrightarrow{\nabla }f(\overrightarrow{r}^{*})=\sum\limits_{i=1}^{M}%
\lambda _{i}\overrightarrow{\nabla }g_{i}(\overrightarrow{r}^{*}),
\label{a30}
\end{equation}
where $\lambda _{1},\lambda _{2},...,\lambda _{M}$ are the Lagrange
multipliers. This represents $N$ sets of equations. We also have constraints
in eq.(\ref{a20}) which are $M$ sets of equations. We have a total of $N+M$
equations and $N+M$ unknowns $(x_{1,}^{*}x_{2}^{*},...,x_{N}^{*})$ and $%
\lambda _{1},\lambda _{2},...,\lambda _{M}$. and therefore they can be
solved in principle. Sometimes this sets of equations are written as
\begin{eqnarray}
&&
\begin{array}{ll}
\partial \Phi /\partial x_{n}=0 & (n=1,2,...,N),
\end{array}
\label{a40} \\
&&
\begin{array}{ll}
\partial \Phi /\partial \lambda _{m}=0 & (m=1,2,...,M).
\end{array}
\label{a50}
\end{eqnarray}
where $\Phi $ is the auxiliary function
\begin{equation}
\Phi (\overrightarrow{r},\lambda _{1},...,\lambda _{M})=f(\overrightarrow{r}%
)-\sum\limits_{i=1}^{M}\lambda _{i}(g_{i}(\overrightarrow{r})-u_{i}).
\label{a60}
\end{equation}

Now we are ready to derive the rule for meaning of the Lagrange multipliers.
The coordinate of the extremum point $\overrightarrow{r}^{*}$ will be a
function of $u_{1},u_{2},...,u_{M}$. Therefore the value of $f(%
\overrightarrow{r})$ at the extremum point $\overrightarrow{r}^{*}$ will be
a function of $u_{1},u_{2},...,u_{M}$ too. We denote this function as $%
F(u_{1},u_{2},...,u_{M}):$%
\begin{equation}
F(u_{1},u_{2},...,u_{M})=f(\overrightarrow{r}^{*}).  \label{a70}
\end{equation}
Now, suppose each $u_{i}$ changed by an infinitesimal amount: $%
u_{i}\rightarrow u_{i}+du_{i}.$ This changes the extremum point $%
\overrightarrow{r}^{*}$ by $d\overrightarrow{r}^{*}$: $\overrightarrow{r}%
^{*}\rightarrow \overrightarrow{r}^{*}+d\overrightarrow{r}^{*}$. The $du_{i}$
and $d\overrightarrow{r}^{*}$ are connected:
\begin{equation}
du_{i}=g_{i}(\overrightarrow{r}^{*}+d\overrightarrow{r}^{*})-g_{i}(%
\overrightarrow{r}^{*})=\overrightarrow{\nabla }g_{i}(\overrightarrow{r}%
^{*})\cdot d\overrightarrow{r}^{*}.  \label{a80}
\end{equation}
The change in $F$ is
\begin{equation}
dF=f(\overrightarrow{r}^{*}+d\overrightarrow{r}^{*})-f(\overrightarrow{r}%
^{*})=\overrightarrow{\nabla }f(\overrightarrow{r}^{*})\cdot d%
\overrightarrow{r}^{*}.  \label{a90}
\end{equation}
The $\overrightarrow{\nabla }f(\overrightarrow{r}^{*})$ can be written from
eq.(\ref{a30}) as
\begin{eqnarray}
dF &=&\sum\limits_{i=1}^{M}\lambda _{i}\cdot \overrightarrow{\nabla }g_{i}(%
\overrightarrow{r}^{*})\cdot d\overrightarrow{r}^{*},  \label{a100} \\
dF &=&\lambda _{1}du_{1}+\lambda _{2}du_{2}+...+\lambda _{M}du_{M}.
\label{a110}
\end{eqnarray}
This clearly tells us that
\begin{equation}
\begin{array}{ll}
\lambda _{i}=\partial F(u_{1},u_{2},...,u_{M})/\partial u_{i} & (i=1,2,...,M)
\end{array}
\label{a120}
\end{equation}
Usually both $F(u_{1},u_{2},...,u_{M})$ and $u_{1},u_{2},...,u_{M}$ have
well defined physical meaning and this relation assigns a physical meaning
to the Lagrange multipliers.

\subsection{Example1. Maximum range of a projectile}

It is instructive to do this elementary exercise using Lagrange multiplier.
If the components of initial velocity are $V_{x},V_{y}$ then the flight time
of the projectile is $2V_{y}/g$ and the range is $%
R(V_{x},V_{y})=2V_{x}V_{y}/g$. The kinetic energy of the projectile is
fixed: $m(V_{x}^{2}+V_{y}^{2})/2=E$. The auxiliary function $\Phi
(V_{x},V_{y},\lambda )$ is
\begin{equation}
\Phi (V_{x},V_{y},\lambda )=2V_{x}V_{y}/g-\lambda \left[
m(V_{x}^{2}+V_{y}^{2})/2-E\right] .  \label{b10}
\end{equation}
Taking partial derivatives with respect to $V_{x},V_{y}$ and $\lambda $ we
get the set of equations
\begin{eqnarray}
2V_{y}/g-\lambda mV_{x} &=&0,  \label{b20} \\
2V_{x}/g-\lambda mV_{y} &=&0,  \label{b30} \\
m(V_{x}^{2}+V_{y}^{2})/2 &=&E.  \label{b40}
\end{eqnarray}
The solutions are
\begin{eqnarray}
V_{x}^{*} &=&V_{y}^{*}=\sqrt{E/m},  \label{b50} \\
\lambda &=&2/mg.  \label{b60}
\end{eqnarray}
The maximum range is
\begin{equation}
R(V_{x}^{*},V_{y}^{*})=(2/mg)E,  \label{b70}
\end{equation}
and $\lambda =dR(V_{x}^{*},V_{y}^{*})/dE$ is satisfied.\textit{\ The meaning
of }$\lambda $\textit{\ here is the rate of increase of maximum range with
energy.}

\subsection{Example2. Microcanonical derivation of quantum gas distributions}

Most applications of Lagrange multipliers involve only one multiplier and
some involve two multipliers. In physics applications involving more than
two multipliers are extremely rare. Here we present a common application in
statistical mechanics involving two multipliers. It is the uninteracting
quantum (Both Bose and fermi) gases.

We follow Huang's statistical mechanics textbook\cite{7} in the following
discussion. We first divide the energy into small intervals $\Delta E_{i}$.
In every interval $\Delta E_{i}$ we have $g_{i}$ single particle levels and $%
n_{i}$ particles occupying these levels. The system size (particle number
and volume) is huge such both the $g_{i}$ and $n_{i}$ are large numbers. But
the energy intervals $\Delta E_{i}$ are small enough such that we can assume
that all $g_{i}$ have the same energy $E_{i}$ where $E_{i}$ is some average
energy value within the $\Delta E_{i}$ interval (say the midpoint of the
interval). By taking a larger and larger system we can make these assumption
more and more accurate.

The number of different ways of putting $n_{i}$ particles into $g_{i}$
degenerate single particle states for Bose particles is
\begin{equation}
w_{i}=\frac{(n_{i}+g_{i}-1)!}{n_{i}!(g_{i}-1)!},  \label{d10}
\end{equation}
and for fermi particles is
\begin{equation}
w_{i}=\frac{g_{i}!}{n_{i}!(n_{i}-g_{i})!}.  \label{d20}
\end{equation}
The total number of ways can be written as
\begin{equation}
W\{n_{i}\}=\prod\limits_{i}w_{i}.  \label{d30}
\end{equation}
We want to maximize $W\{n_{i}\}$ subject to the constraints
\begin{eqnarray}
\sum\limits_{i}n_{i} &=&N,  \label{d40} \\
\sum\limits_{i}n_{i}E_{i} &=&U.  \label{d50}
\end{eqnarray}
Instead of maximizing $W\{n_{i}\}$ we maximize $k_{B}\ln (W\{n_{i}\})$ for
convenience. We write the auxiliary function $\Phi (\{n_{i}\},\lambda )$
\begin{equation}
\Phi (\{n_{i}\},\lambda )=k_{B}\sum\limits_{i}\ln (w_{i})-\lambda _{1}\left(
\sum\limits_{i}n_{i}E_{i}-U\right) -\lambda _{2}\left(
\sum\limits_{i}n_{i}-N\right) ,  \label{d60}
\end{equation}
and use the Stirling approximation to express the factorials in $\ln w_{i}$
and set derivatives $\partial \Phi /\partial n_{i}$ to zero to obtain the
equations that $\{n_{i}\}$ satisfy. When they are solved we obtain
\begin{equation}
n_{i}^{*}=\frac{g_{i}}{\exp \left( \lambda _{1}E_{i}/k_{B}+\lambda
_{2}/k_{B}\right) \pm 1},  \label{d70}
\end{equation}
where $+$ is for fermions and $-$ is for bosons in the denominator. Here $%
g_{i}>>1$ is used to obtain this result. The $\lambda _{1}$ and $\lambda
_{2} $ is found by solving the equations $\sum\limits_{i}n_{i}^{*}=N$ and $%
\sum\limits_{i}n_{i}^{*}E_{i}=U$ simultaneously.

Now we have the problem of identifying the thermodynamical meaning of $%
\lambda _{1}$ and $\lambda _{2}$. Staying within the microcanonical ensemble
this identification requires considerable amount of supplementary argument.
Of course one can derive eq. (\ref{d70}) using grand canonical ensemble
where the quantities appearing in place of $\lambda _{1}$ and $\lambda _{2}$
have fixed meanings in the grand ensemble. But in teaching statistical
mechanics we usually want to show that all ensembles are equivalent an we
should be able to find physical meaning of $\lambda _{1}$ and $\lambda _{2}$
staying in microcanonical ensemble too. Our rule makes this identification a
trivial exercise as shown below.

In the microcanonical ensemble the entropy is
\begin{equation}
S=k_{B}\ln \left( \sum\limits_{\{n_{i}\}}W\{n_{i}\}\right) ,  \label{d80}
\end{equation}
where the sum runs over all possible $\{n_{i}\}\,$sets. It is shown in
statistical mechanics textbooks that in the thermodynamic limit this is
equal to $k_{B}\ln W\{n_{i}^{*}\}$. Therefore $k_{B}\ln W\{n_{i}^{*}\}\,$is
the entropy $S(U,N)$ as a function of internal energy $U$ and the particle
number $N$. From the rule it follows that the $\lambda _{1}$ and $\lambda
_{2}\,$are
\begin{eqnarray}
\lambda _{1} &=&\left( \partial S/\partial U\right) _{N}  \label{d90} \\
\lambda _{2} &=&(\partial S/\partial N)_{U}.  \label{d100}
\end{eqnarray}
The first one is easily interpreted as $1/T\,$where $T$ is the temperature
of the system. In order to interpret the second one we use the well-known
relation
\begin{equation}
(\partial S/\partial N)_{U}\cdot (\partial N/\partial U)_{S}\cdot (\partial
U/\partial S)_{N}=-1.  \label{d110}
\end{equation}
From thermodynamics $(\partial U/\partial S)_{N}=T$ and $(\partial
N/\partial U)_{S}=\mu ^{-1}$ where $\mu $ is the chemical potential of the
system. Then we obtain $\lambda _{2}=(\partial S/\partial N)_{U}=-\mu /T$.

As can be seen from this example, in statistical mechanical applications the
rule makes supplementary arguments to interpret multipliers unnecessary. I
invite the reader to try to find another argument as short as this one
(staying within microcanonical ensemble) without using the rule. In
statistical mechanics (unlike quantum mechanics for example) both
constrained and optimized quantities always have well defined physical
meanings and identification of thermodynamical meaning of the multipliers is
only one-line argument using this rule. Once the rule is introduced
beforehand (in an appendix of a book or as a handout in a class for example)
identification of multipliers will be  a very precise and concise argument.

\section{Derivation of the rule in variational problems and examples}

We will not discuss the theory of variational calculus here or how the
Lagrange multipliers are used in variational calculus. We will merely state
basic results and derive the equivalent formula for the interpretation of
Lagrange multipliers.

In the simplest case of variational problems we are trying to find the
function $y(x)$ that makes the integral
\begin{equation}
I[y]=\int\limits_{a}^{b}f(x,y,y^{\prime })dx,  \label{v10}
\end{equation}
an extremum where $y^{\prime }=dy/dx$ and $y(a)$ and $y(b)$ are fixed. The
function $y^{*}$ that makes $I[y]$ an extremum satisfies the Euler
differential equation
\begin{equation}
\frac{\delta I[y^{*}]}{\delta y^{*}}=\frac{\partial f}{\partial y^{*}}-\frac{%
d}{dx}\left( \frac{\partial f}{\partial (y^{*})^{\prime }}\right) =0.
\label{v20}
\end{equation}
If we have constraints
\begin{equation}
\begin{array}{ll}
J_{i}[y]=\int\limits_{a}^{b}g_{i}(x,y,y^{\prime })dx=U_{i} & (i=1,2,...,M),
\end{array}
\label{v30}
\end{equation}
then we solve the differential equation
\begin{equation}
\frac{\delta I[y^{*}]}{\delta y^{*}}=\sum\limits_{i=1}^{M}\lambda _{i}\frac{%
\delta J_{i}[y^{*}]}{\delta y^{*}}  \label{v40}
\end{equation}
for $y^{*}(x)$ which depends on Lagrange multipliers $\lambda
_{1},...,\lambda _{M}$ parametrically. When we put $y^{*}$ in the constraint
equations eq.(\ref{v30}) we obtain a set of algebraic (in general nonlinear)
sets of equation for $\lambda _{1},...,\lambda _{M}$. The solutions for $%
\lambda _{1},...,\lambda _{M}$ will be a function of the parameters $%
U_{1},U_{2},...,U_{M}$ and when the multipliers are put in $y^{*}(x)$ back,
the $y^{*}(x)$ itself will be a function of $U_{1},U_{2},...,U_{M}$
parameters. When we feed $y^{*}(x)\,$in the integral for $I[y]$ we get $%
I[y^{*}]$ which will be a function of $U_{1},U_{2},...,U_{M}$ too. We denote
this function as $\mathrm{I}(\mathbf{U})$ where $\mathbf{U\,}$stands for $%
U_{1},U_{2},...,U_{M}$. So we have $\mathrm{I}(\mathbf{U})=I[y^{*}]$.

Now we change the parameters $U_{1},U_{2},...,U_{M}$ by an infinitesimal
amount: $U_{i}\rightarrow U_{i}+dU_{i}$ and this changes the solution $y^{*}$
by an infinitesimal amount: $y^{*}\rightarrow y^{*}+\Delta y^{*}$ where $%
\Delta y^{*}\,$is an infinitesimal function. this changes $\mathrm{I}(%
\mathbf{U})\,$by
\begin{equation}
d\mathrm{I}(\mathbf{U})=I[y^{*}+\Delta y^{*}]-I[y^{*}]=\int\limits_{a}^{b}%
\frac{\delta I[y^{*}]}{\delta y}\left( \Delta y^{*}\right) dx.  \label{v50}
\end{equation}
For the constraint we also have
\begin{equation}
\begin{array}{ll}
dU_{i}=J_{i}[y^{*}+\Delta y^{*}]-J_{i}[y^{*}]=\int\limits_{a}^{b}\left(
\delta J_{i}[y^{*}]/\delta y\right) \left( \Delta y^{*}\right) dx. &
(i=1,2,...,M).
\end{array}
\label{v60}
\end{equation}
Putting the $\delta I[y^{*}]/\delta y$ from eq.(\ref{v40}) into the eq.(\ref
{v50}) and comparing to the eq.(\ref{v60}) we obtain
\begin{equation}
d\mathrm{I}(\mathbf{U})=\lambda _{1}dU_{1}+\lambda _{2}dU_{2}+...,\lambda
_{M}dU_{M},  \label{v70}
\end{equation}
which tells us that
\begin{equation}
\begin{array}{ll}
\lambda _{i}=\partial \mathrm{I}(U_{1},U_{2},...,U_{M})/\partial U_{i} &
(i=1,2,...,M).
\end{array}
\label{v90}
\end{equation}
Usually both $\mathrm{I}(\mathbf{U})$ and $\mathbf{U}$ have physical meaning
and this prescription assigns a physical meaning to the Lagrange multipliers
$\lambda _{1},...,\lambda _{M}$.

\subsection{Example1. Hanging chain}

In mechanics Lagrange multipliers are used in many places. One kind of
application involves constrained motion in Lagrangian formulation of
mechanics.\cite{Goldstein} In this application Lagrange multipliers are time
dependent and it is well known that they are equal to constraint forces. The
author does not see how our rule applies to this case if it does. But it is
possible to find mechanics problems where the rule applicable. Here we give
an example to this.

Suppose a chain of fixed length $L$ and density $\rho $ is hanging from two
points $(\pm a,0)$ in the $x-y$ plane. The shape of the chain $y(x)$ will be
such that the potential energy integral
\begin{equation}
V[y]=\rho g\int_{-a}^{a}y\sqrt{1+(y^{\prime })^{2}}dx  \label{av10}
\end{equation}
will be a minimum under the constraint that the length of the chain is
fixed:
\begin{equation}
\int_{-a}^{a}\sqrt{1+(y^{\prime })^{2}}dx=L  \label{av20}
\end{equation}
Setting up the Euler equation with a Lagrange multiplier is straightforward.
Solving the arising differential equation and imposing the boundary
conditions $y(\pm a)=0$ we obtain the solution
\begin{eqnarray}
y^{*}(x) &=&\frac{1}{\alpha }\left[ \cosh (\alpha x)-\cosh (\alpha a)\right]
,  \label{av30} \\
\lambda &=&-\frac{\rho g}{\alpha }\cosh (\alpha a),  \label{av40}
\end{eqnarray}
where $\alpha $ is an integration constant. To obtain a relation between $%
\alpha $ and $L$ we put $y^{*}(x)$ in the eq.(\ref{av20}) to obtain
\begin{equation}
\frac{2}{\alpha }\sinh (\alpha a)=L.  \label{av50}
\end{equation}
We also calculate $V[y^{*}]$ by putting the $y^{*}$ in the eq.(\ref{av10})
to obtain
\begin{equation}
V[y^{*}]=\rho g\left( \frac{a}{\alpha }-\frac{\sinh (2\alpha a)}{\alpha ^{2}}%
\right) .  \label{av60}
\end{equation}
To show that $dV[y^{*}]/dL=\lambda $ we write $dV[y^{*}]/dL$ as
\begin{equation}
\frac{dV[y^{*}]}{dL}=\frac{dV[y^{*}]/d\alpha }{dL/d\alpha },  \label{av70}
\end{equation}
and calculate $dL/d\alpha $ and $dV[y^{*}]/d\alpha $ from eqs. (\ref{av50},%
\ref{av60}) to obtain
\begin{equation}
\frac{dV[y^{*}]}{dL}=-\frac{\rho g}{\alpha }\cosh (\alpha a)=\lambda .
\label{av80}
\end{equation}
We showed that $dV[y^{*}]/dL=\lambda $ holds.

Now what is the physical meaning of the Lagrange multiplier $\lambda $?
Imagine pulling the chain from one end by an infinitesimal amount $dL$. The
work done is $T\cdot dL$ where $T$ is the tension at the endpoints and this
should be equal to the potential energy difference $%
V[y^{*}(L-dL)]-V[y^{*}(L)]=-dV[y^{*}]$. \textit{Therefore }$%
T=-dV[y^{*}]/dL=-\lambda $\textit{\ and we have the physical interpretation
that }$-\lambda $\textit{\ is equal to the tension at the ends of the chain.}
As usually the case in mechanics, here too the multiplier turns out to be a
constraint force.

Here application of the rule $dV[y^{*}]/dL=\lambda $ makes the
interpretation of multiplier a simple one-line argument. I invite the reader
to try to find another argument as short as this one for the physical
meaning of the multiplier without using the rule.

\subsection{Example2. Constrained brachistochrone problem}

Here we present a mechanics problem that the multiplier is not a constraint
force and yet has a very precise meaning. We have a rail that extends from
point $A$ $(x=0,y=h)$ to the point $B$ $(x=a,y=0)$ and the rail has a fixed
length $L\geq \sqrt{a^{2}+h^{2}}$. A mass slides on it from point A to point
B. What is the curve that delivers the shortest travel time?

Here the travel time
\begin{equation}
T[y]=\sqrt{\frac{1}{2g}}\int_{0}^{a}\sqrt{\frac{1+(dy/dx)^{2}}{h-y}}dx,
\label{e10}
\end{equation}
is minimized with the constraint
\begin{equation}
U[y]=\int_{0}^{a}\sqrt{1+(dy/dx)^{2}}dx=L.  \label{e20}
\end{equation}
The Euler equation
\begin{equation}
\frac{\delta T[y]}{\delta y}=\lambda \frac{\delta U[y]}{\delta y},
\label{e30}
\end{equation}
yields (after integrating once) the differential equation
\begin{equation}
C\sqrt{1+(dy/dx)^{2}}=\frac{1}{\sqrt{2g(h-y)}}-\lambda  \label{e40}
\end{equation}
where $C$ is an integration constant. Because of difficulties with
evaluating some integrals we were not able to solve this problem in closed
form. Therefore we do not give a explicit demonstration that the rule $%
\lambda =dT[y^{*}]/dL$ holds. But assuming that the multiplier is calculated
(numerically or using some approximation) we can tell physical meaning of
it. \textit{It is rate of increase of optimal travel time with increasing
length }$L$. Physically we expect that as $L$ starts increasing from the
initial value $\sqrt{a^{2}+h^{2}}$ the travel time $T[y^{*}]$ should
decrease (hence negative $\lambda $) until an optimal length $L_{O}$ is
reached. When $L=L_{O}$ we have $\lambda =0$ and problem reduces to
unconstrained brachistochrone problem and travel time is the absolute
minimum. As $L$ is increased further the travel time should increasing again
yielding positive $\lambda $. Here we see that, using both our knowledge of
physics and the rule enables us to find the sign of $\lambda $ without
solving the problem.

\subsection{Example3. Minimum uncertainty wave packet}

In some quantum problems the Lagrange multipliers have no apparent physical
meanings. These cases arises especially for normalization and overlap
integral constraints. Here we give an example that has two Lagrange
multipliers. One of the Lagrange multipliers has a clear physical meaning
obeys the rule and the other does not.

We want to find the wave packet with the minimum momentum uncertainty
\begin{equation}
M[\psi ]=\int\limits_{-\infty }^{\infty }\psi ^{*}\widehat{P}^{2}\psi
dx-\left( \int\limits_{-\infty }^{\infty }\psi ^{*}\widehat{P}\psi dx\right)
^{2}=\sigma _{P}^{2}  \label{cv10}
\end{equation}
with the conditions that position uncertainty is fixed
\begin{equation}
K[\psi ]=\int\limits_{-\infty }^{\infty }\psi ^{*}x^{2}\psi dx-\left(
\int\limits_{-\infty }^{\infty }\psi ^{*}x\psi dx\right) ^{2}=\sigma
_{x}^{2},  \label{cv20}
\end{equation}
and the wave function is normalized
\begin{equation}
N[\psi ]=\int\limits_{-\infty }^{\infty }\psi ^{*}\psi dx=u.  \label{cv25}
\end{equation}
Here $\widehat{P}=-i\hbar d/dx$ momentum operator and for the sake of the
argument we consider normalization integral with a general fixed value $u$.
We later take $u=1$ after showing that $u\neq 1$ has no solution. Treating $%
\psi $ and $\psi ^{*}$ as independent variables the Euler equation with two
Lagrange multipliers
\begin{equation}
\frac{\delta M[\psi ]}{\delta \psi ^{*}}=\lambda _{1}\frac{\delta K[\psi ]}{%
\delta \psi ^{*}}+\lambda _{2}\frac{\delta N[\psi ]}{\delta \psi ^{*}}
\label{cv30}
\end{equation}
yields the equation
\begin{equation}
-\hbar ^{2}\psi ^{\prime \prime }(x)+2i\hbar b\psi ^{\prime }(x)-\lambda
_{1}\left( x^{2}-2ax\right) \psi (x)-\lambda _{2}\psi (x)=0  \label{cv35}
\end{equation}
where $a$ and $b$ are the parameters defined as
\begin{eqnarray}
a &=&\int\limits_{-\infty }^{\infty }\psi (x)^{*}x\psi (x)dx,  \label{cv40}
\\
b &=&\int\limits_{-\infty }^{\infty }\psi ^{*}(x)\widehat{P}\psi (x)dx.
\label{cv50}
\end{eqnarray}
As a consistency condition the solution of eq.(\ref{cv35}) should satisfy
eqs.(\ref{cv40},\ref{cv50}). Making the transformation $\psi
(x)=e^{ibx/\hbar }\varphi (x)$ we obtain the equation
\begin{equation}
-\hbar ^{2}\varphi ^{\prime \prime }(x)-\lambda _{1}\left( x-a\right)
^{2}\varphi (x)=\left( \lambda _{2}+b^{2}-\lambda _{1}a^{2}\right) \varphi .
\label{cv60}
\end{equation}
This is the Schrodinger equation for a Harmonic oscillator problem ($m=1/2$
and $w=\sqrt{-4\lambda _{1}}$) and normalizable solutions are possible only
for $\lambda _{1}<0$. The solutions are the harmonic oscillator
eigenfunctions $\phi _{0},\phi _{1},...,\phi _{n},...$ where $\phi _{n}(x)\,$%
is normalized in the usual way: $\int \left| \phi _{n}(x)\right| ^{2}dx=1$.
But the $\psi (x)$ must satisfy the normalization condition given in eq.(\ref
{cv25}). Therefore we take $\psi _{n}(x)=\sqrt{u}e^{ibx/\hbar }\phi _{n}(x)$%
. With this $\psi (x)$ we get the following expectation values
\begin{eqnarray}
\int\limits_{-\infty }^{\infty }\psi _{n}^{*}(x)x\psi _{n}(x)dx &=&ua,
\label{cv64} \\
\int\limits_{-\infty }^{\infty }\psi _{n}^{*}(x)\widehat{P}\psi _{n}(x)dx
&=&ub.  \label{cv65}
\end{eqnarray}
The solution does not satisfy the consistency conditions given in eqs.(\ref
{cv40},\ref{cv50}). Therefore consistent solutions are possible only for $%
u=1 $. We take $u=1$ for the rest of the discussion.

All $\psi _{n}(x)$ are stationary solutions and they all have $\langle
\widehat{P}\rangle =b$. For minimum momentum uncertainty we must have
minimum $\langle \widehat{P}^{2}\rangle $. The $\langle \psi _{n}\left|
\widehat{P}^{2}\right| \psi _{n}\rangle $ is easily calculated as $%
b^{2}+(n+1/2)\hbar w/2$. Therefore for minimum momentum uncertainty we must
take the ground state:
\begin{equation}
\phi (x)=\left( \frac{2k}{\pi }\right) ^{1/4}e^{-k(x-a)^{2}},  \label{cv80}
\end{equation}
where $\langle x\rangle =a.$ Putting this back into the eq.(\ref{cv60})
yields
\begin{eqnarray}
\lambda _{1} &=&-4k^{2}\hbar ^{2}  \label{cv90} \\
2k\hbar ^{2} &=&\lambda _{2}+b^{2}-\lambda _{1}a^{2}.  \label{cv95}
\end{eqnarray}
We also calculate $\langle x^{2}\rangle $ and $\langle \widehat{P}%
^{2}\rangle $ for $\psi _{0}(x)$ as
\begin{eqnarray}
\langle x^{2}\rangle &=&a^{2}+1/(4k)  \label{cv120} \\
\langle \widehat{P}^{2}\rangle &=&b^{2}+\hbar ^{2}k.  \label{cv130}
\end{eqnarray}
From this $\sigma _{x}^{2}=\langle x^{2}\rangle -\langle x\rangle
^{2}=1/(4k) $ and ($\sigma _{P}^{2})_{\min }=\langle \widehat{P}^{2}\rangle
-\langle \widehat{P}\rangle ^{2}=\hbar ^{2}k$ are obtained. This yields the
uncertainty principle ($\sigma _{P}^{2})_{\min }\sigma _{x}^{2}=\hbar ^{2}/4$%
. The Lagrange multiplier $\lambda _{1}$ has the value $\lambda _{1}=-\hbar
^{2}/4(\sigma _{x}^{2})^{2}$ and satisfies the rule
\begin{equation}
\lambda _{1}=\frac{d(\sigma _{P}^{2})_{\min }}{d\sigma _{x}^{2}}=-\frac{%
\hbar ^{2}}{4(\sigma _{x}^{2})^{2}}=-\frac{(\sigma _{P}^{2})_{\min }}{\sigma
_{x}^{2}}.  \label{cv140}
\end{equation}
\textit{The physical meaning of the lagrange multiplier is the rate of
change of minimum momentum uncertainty with position uncertainty which is
equal to negative ratio of both uncertainties.}

The Lagrange multiplier $\lambda _{2}$ has the value $2k\hbar
^{2}-b^{2}+\lambda _{1}a^{2}$ and it has no obvious physical interpretation.
We cannot apply the rule $\lambda _{2}=d(\sigma _{P}^{2})_{\min }/du$ either
because we showed that the solution does not exist for $u\neq 1$. Even if it
existed it would still be difficult to attach a meaning to it because the
for $u\neq 1$ we have no physical meaning to assign to the variable $u$.

\end{document}